\documentclass[preprint1]{aastex}

\usepackage{amsmath}

\slugcomment{}

\newcommand{\be}{\begin{equation}}
\newcommand{\ee}{\end{equation}}

\shorttitle{$H(z)$ in the $R_{\rm h}=ct$ Universe}
\shortauthors{Melia \& McClintock}

\begin{document}

\title{A Test of Cosmological Models using high-$z$ Measurements of $H(z)$}
\author{Fulvio Melia\altaffilmark{1,\ddag} and Thomas M. McClintock\altaffilmark{2}}
\altaffiltext{1}{Department of Physics, The Applied Math Program, and Department of Astronomy,
The University of Arizona, AZ 85721, USA; fmelia@email.arizona.edu.}
\altaffiltext{2}{Department of Physics, The University of Arizona, AZ 85721, USA; tmcclintock89@gmail.com}
\altaffiltext{\ddag}{John Woodruff Simpson Fellow.}

\begin{abstract}
The recently constructed Hubble diagram using a combined sample of
SNLS and SDSS-II Type Ia SNe, and an application of the Alcock-Paczy\'nski
(AP) test using model-independent Baryon Acoustic Oscillation data, have 
suggested that the principal constraint underlying the cosmic expansion
is the total equation-of-state of the cosmic fluid, rather than that of
its dark energy. These studies have focused on the critical redshift
range ($0\lesssim z\lesssim 2$) within which the transition from decelerated
to accelerated expansion is thought to have occurred, and they suggest
that the cosmic fluid has zero active mass, consistent with a constant
expansion rate. The evident impact of this conclusion on cosmological
theory calls for an independent confirmation. In this paper, we
carry out this crucial one-on-one comparison between the $R_{\rm h}=ct$
Universe (an FRW cosmology with zero active mass) and $w$CDM/$\Lambda$CDM, using
the latest high-$z$ measurements of $H(z)$. Whereas the Type Ia SNe
yield the integrated luminosity distance, while the AP diagnostic 
tests the geometry of the Universe, the Hubble parameter directly
samples the expansion rate itself. We find that the model-independent
cosmic chronometer data prefer $R_{\rm h}=ct$ over $w$CDM/$\Lambda$CDM with a
BIC likelihood of $\sim 95\%$ versus only $\sim 5\%$, in strong support 
of the earlier SNeIa and AP results. This contrasts with a recent
analysis of $H(z)$ data based solely on BAO 
measurements which, however, strongly depend on the assumed 
cosmology. We discuss why the latter approach is inappropriate for model
comparisons, and emphasize again the need for truly model-independent 
observations to be used in cosmological tests.

\end{abstract}

\keywords{cosmology: cosmological parameters -- cosmology: distance scale -- cosmology: observations -- 
cosmology: theory -- galaxies}

\section{Introduction}
Modern cosmology is based on the Friedmann-Robertson-Walker (FRW) metric to describe
the spacetime expansion of the cosmic fluid, often assumed to be shear free (i.e.,
`perfect'), and comprised of at least three components. Two of these,
matter ($\rho_{\rm m}$) and radiation ($\rho_{\rm r}$), are readily observed, while 
the third, a poorly understood `dark energy' ($\rho_{\rm de}$),  is inferred 
from the analysis of the distance-redshift relationship in Type Ia SNe 
(Perlmutter et al. 1998; Riess et al. 1998; Schmidt et al. 1998). The standard model 
of cosmology, which we refer to as $\Lambda$CDM when $\rho_{\rm de}$ is a cosmological
constant with equation-of-state $w_{\rm de}\equiv p_{\rm de}/\rho_{\rm de}=-1$, and $w$CDM
when it is not, adopts these elements, but the poorly known $w_{\rm de}$ burdens the 
theory with several unknown parameters that need to be optimized while fitting the data.

In spite of this limitation, $w$CDM/$\Lambda$CDM has done remarkably well in accounting for
the observations, though the ever improving precision of the various measurements
is beginning to uncover tension---in some cases, actual inconsistencies---with the
model predictions. In fact, the most recent tests of the standard model have revealed
that the principal constraint underlying the cosmic expansion appears to be the total 
equation-of-state for the cosmic fluid rather than that of its dark energy. The 
observational evidence now favors a cosmic fluid with zero active mass, i.e., 
$\rho+3p=0$, where $\rho\equiv \rho_{\rm m}+\rho_{\rm r}+\rho_{\rm de}$ and $p\equiv  
p_{\rm m}+p_{\rm r}+p_{\rm de}$ are, respectively, its total energy density 
and pressure (see also Melia 2015b). This result, however, would exclude a cosmological 
constant as a representation of dark energy.

It is therefore critically important to seriously examine the zero active mass
condition with the highest precision permitted by current observations.
In the literature, the FRW cosmology with zero 
active mass is known as the $R_{\rm h}=ct$ Universe (Melia 2007; Melia 
and Shevchuk 2012). This model has been compared with $w$CDM/$\Lambda$CDM 
(an FRW cosmology lacking this condition) in many
one-on-one tests, at both low and high redshifts (see, e.g., Melia and Maier
2013; Wei et al. 2013, 2014a, 2014b, 2015a; Melia et al. 2015a). Thus far, 
$R_{\rm  h}=ct$ has been favored by the data, with model selection tools 
typically yielding likelihoods of $\sim 90\%$ versus only $\sim 10\%$ for
$w$CDM/$\Lambda$CDM. However, these results have not yet been universally
accepted, and several counterclaims have emerged in recent years. These may
be loosely grouped together into four primary categories: (1) that $R_{\rm h}$ 
(representing the gravitational radius and, therefore, also the Hubble radius)
does not really have any physical meaning or bearing on the observations; 
this claim has been made chiefly by van Oirschot et al. (2010), Bilicki 
\& Seikel (2012), and Lewis \& van Oirschot (2012); (2) that the zero active
mass condition $\rho+3p=0$ cannot be made consistent with the actual
constitutents in the cosmic fluid (Lewis 2013); (3) that the measurements of
$H(z)$ as a function of redshift (the primary topic of the present paper)
favor the concordance model over $R_{\rm h}=ct$ (Bilicki \& Seikel 2012;
Shafer 2015); and (4) that the analysis of Type Ia SNe also favor the
concordance model over $R_{\rm h}=ct$ (Bilicki \& Seikel 2012; Shafer
2015). These papers, and those published in response to them (see, e.g.,
Melia \& Maier 2013; Bikwa et al. 2012; Melia \& Shevchuk 2012; Melia 
2012, 2014, 2015a), have generated a very important discussion that 
we aim to continue here. Specifically, in \S~3 below, we will describe 
at length why the choice of truly model-independent data, and their
analysis using sound statistical practices, is of utmost importance
to any serious attempt at comparing different cosmological models in 
an unbiased fashion.

Of the previous tests favoring $R_{\rm h}=ct$ over $w$CDM/$\Lambda$CDM,
two are particularly noteworthy. In the first of these, a combined sample 
of 613 supernova events from SNLS~(Guy et al. 2010) and the newly released 
SDSS\nobreakdash-II~(Sako et al. 2015) was used to show that in an unbiased 
pairwise comparison, the Bayes Information Criterion (BIC) favors the $R_{\rm h}=ct$ 
Universe with a likelihood of $\approx 88\%$ versus $\approx 12\%$ for the standard 
model (Wei et al. 2015b; Melia et al. 2015b). This combined sample spans the 
critical redshift range ($0\lesssim z\lesssim 1.2$) in which the presumed transition 
from deceleration to acceleration is thought to have occurred. This outcome 
reaffirms the influence of dark energy, but calls into question the widely held 
belief that the Universe is currently accelerating. In other words, when a neutral
calibration of the SN data is used to compare models, the evidence does not
appear to support the interpretation of dark energy as a cosmological constant.

In a second test, an application of the Alcock-Paczy\'nski (AP) diagnostic 
using the model-independent anisotropic distribution of Baryon Acoustic
Oscillation (BAO) peaks at average redshifts $\langle z\rangle=0.57$ and 
$\langle z\rangle=2.34$ has shown that the BAO data disfavor $w$CDM/$\Lambda$CDM 
at better than $2.7\sigma$ (Melia and L\'opez-Corredoira 2015). In the 
context of expanding FRW cosmologies, these measurements instead strongly 
prefer the zero active mass equation-of-state, confirming the conclusion drawn 
earlier from the Type Ia SN Hubble diagram.

Each of these results is quite significant in its own right; together, they
make quite a compelling case. But given their evident impact, they need
to be confirmed using another independent diagnostic
in the critical redshift range sampled by the SNeIa and BAO observations. 
In this paper, we focus on several new measurements of the Hubble parameter
$H(z)$ within the redshift range $0\lesssim z\lesssim 2.5$, using luminous red 
galaxies as cosmic chronometers (Moresco 2015). These measurements were 
made without priors, and are therefore model independent---ideally suited 
to a one-on-one comparison between $w$CDM/$\Lambda$CDM and $R_{\rm h}=ct$. 

With the latest extension of the $H(z)$ coverage to $z\sim2$, the cosmic 
chronometer data are now fully compatible with the SNeIa and BAO measurements. 
But whereas the SN comparison is based on the (integrated) luminosity distance, 
and the AP diagnostic tests the predicted geometry, the $H(z)$ measurements 
directly probe the expansion rate itself. Thus, although they sample the same 
redshift range, these three diagnostics test the cosmologies in distinctly 
different ways, and each therefore contributes to the self-consistency of the 
tests in its own unique way. In this paper, we carry out the critical 
one-on-one comparison between $R_{\rm h}=ct$ and $w$CDM/$\Lambda$CDM using the 
latest cosmic chronometer data. In so doing, we demonstrate that this 
independent test also strongly favors $R_{\rm h}=ct$ over $w$CDM/$\Lambda$CDM, 
thereby strengthening the argument in favor of the zero active mass 
condition.

\section{The Cosmic Chronometers}
Since the Hubble rate depends on the expansion parameter $a(t)$
according to 
\begin{equation}
H(z) = {\dot{a}\over a}\;,
\end{equation}
and $(1+z)=a_0/a$, where $a_0$ is the expansion parameter today, $H(z)$ may
be measured directly from the time-redshift derivative $dt/dz$ using
\begin{equation}
H(z) = -{1\over 1+z}{dz\over dt}\;.
\end{equation}
Measuring $H(z)$ using the differential age of the Universe therefore
circumvents the limitations associated with the use if integrated
histories. 

\begin{figure}
\begin{center}
\includegraphics[width=1.0\linewidth]{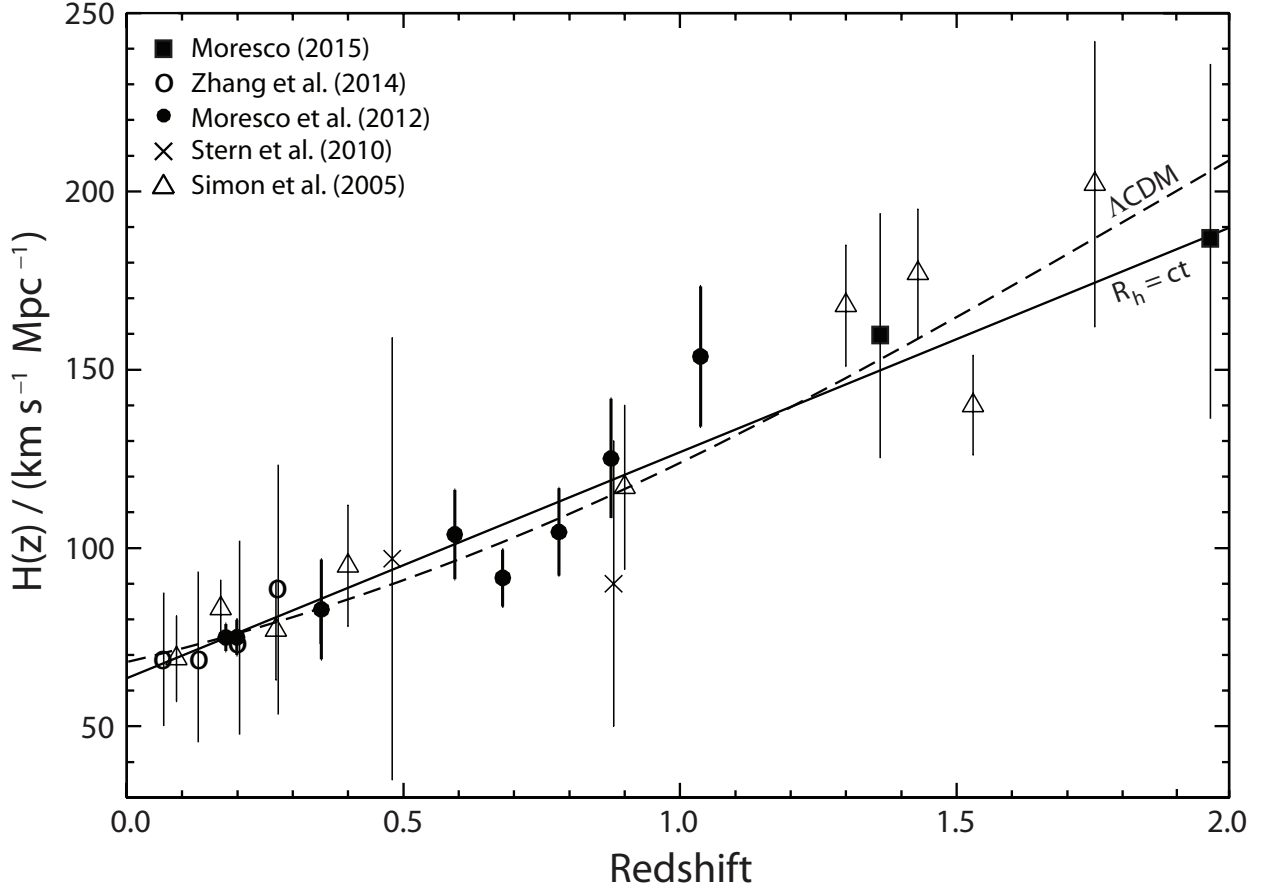}
\end{center}
\caption{Twenty five (model-independent) measurements of $H(z)$, with error bars, 
and a comparison with two theoretical models:
{\it (dashed)} the optimized $w$CDM cosmology, with best-fit parameters $H_0=68.0\pm
7.9$ km s$^{-1}$ Mpc$^{-1}$, $\Omega_{\rm m}=0.31\pm 0.16$, and $w_{\rm de}=-0.91_{+0.42}^{-0.94}$;
and {\it (solid)} the $R_{\rm h}=ct$ Universe, with its sole parameter $H_0=63.3\pm7.7$ km s$^{-1}$
Mpc$^{-1}$. The reduced $\chi^2_{\rm dof}$ (24 degrees of freedom) for $R_{\rm h}=ct$ is
0.56. The corresponding value for $w$CDM (22 degrees of freedom) is $\chi^2_{\rm dof}=0.58$.
The BIC favors $R_{\rm h}=ct$ over $w$CDM with a
likelihood of $\sim 94.3\%$ versus $\sim 5.7\%$.}
\vskip 0.2in
\end{figure}

Galaxies evolving passively on a time scale much longer
than their age difference are among the best cosmic chronometers one
may use for this purpose (Jimenez and Loeb 2002). Less than 1 per cent
of the stellar mass in the most massive galaxies formed at $z<1$
(Dunlop et al. 1996; Spinrad et al. 1997; Cowie et al. 1999;
Heavens et al. 2004; Thomas et al. 2005; Panter et al. 2007).
In fact, star formation ceased by redshift $z\sim 3$
(Thomas et al. 2005) in galaxy clusters, while quite generally,
all systems with stellar mass over $\sim 5\times 10^{11}\;M_\odot$
ended their star formation activity by $z\sim 2$ (Treu et al. 2005).

It is therefore reasonable to assume that galaxies in the highest
density regions of clusters have been evolving passively since
$z\sim 2$, tracing the so-called `red envelope,' which samples the
oldest stars in the Universe at every redshift. For this reason,
these structures have been used as reliable cosmic chronometers 
in several extended studies (Stern et al. 2010a,b; Moresco et al. 
2012a,b), culminating with the most recent measurements
at $z\sim 2$ (Moresco 2015).

The data shown in figure~1 were assembled from the compilations
of Simon et al. (2005), Stern et al. (2010), Moresco et al. (2012a)
and Moresco (2015). We emphasize the fact that all of these
measurements are model-independent. Other kinds of measurement of 
$H(z)$, based on the identification of BAO and the Alcock-Paczy\'nski
distortion from galaxy clustering, depend on how `standard rulers'
evolve with redshift, rather than how cosmic time changes with $z$.
Unfortunately, the values of $H(z)$ measured in these different
ways are sometimes combined to produce an overall $H(z)$ versus
$z$ diagram, but such compilations cannot be used to test
different models because the second approach necessarily adopts
a particular cosmology and is therefore model-dependent (Blake
et al. 2012). Only the cosmic chronometer measurements are
truly model-independent and therefore suitable for comparing
$w$CDM/$\Lambda$CDM to other models, such as $R_{\rm h}=ct$.

\section{Discussion}
In the $R_{\rm h}=ct$ Universe, $a(t)\propto t$, so the Hubble
parameter scales very simply with redshift:
\begin{equation}
H(z) = H_0 (1+z)\;.
\end{equation}
Since $H_0$ is the sole free parameter in this model, the corresponding
theoretical curve in figure~1 can be adjusted only by sliding it 
vertically. There is no freedom to alter its gradient with redshift. 
Because the redshift coverage with these measurements now extends to 
$z\sim 2$, this particular test is therefore very constraining on models, 
such as $R_{\rm h}=ct$, that do not predict a transition from deceleration 
to acceleration somewhere near $z\sim 1$. 

$w$CDM/$\Lambda$CDM is characterized by a larger number of free parameters
because the properties of dark energy are so poorly known. At the
very least, one must include the Hubble constant $H_0$, the fractional 
energy density $\Omega_{\rm m}$ in the form of matter ($\equiv 
\rho_{\rm m}[t_0]/\rho_{\rm c}$, where $\rho_{\rm c}\equiv 3c^2H_0^2/8\pi G$ 
is today's critical density), the dark energy equation-of-state parameter, 
$w_{\rm de}\equiv p_{\rm de}/\rho_{\rm de}$, and the spatial curvature
constant $k$. But to keep the comparison as simple and favorable to 
$w$CDM/$\Lambda$CDM as possible, we will here adopt the prior value $k=0$ 
in the case of $w$CDM, and $w_{\rm de}=-1$ for $\Lambda$CDM, thus reducing 
the total number of adjustable parameters to an essential three in each case. 
The Hubble parameter in this cosmology is therefore given by the expression
\begin{equation}
H(z)=H_0\left[\Omega_{\rm m}(1+z)^3+\Omega_{\rm r}(1+z)^4+\Omega_{\rm de}
(1+z)^{3(1+w_{\rm de})}\right]^{1/2}\;.
\end{equation}
Here, $\Omega_{\rm r}$ for radiation and $\Omega_{\rm de}$ for dark energy
are defined analogously to $\Omega_{\rm m}$. The scaling of each density
with redshift assumes that all three components evolve independently of
each other, which may be reasonable in the local Universe (i.e., at least
out to $z\sim 2$). Note that although we formally include $\Omega_{\rm r}$ 
in this expression, practically speaking its value ($\sim 6\times 10^{-5}$)
remains too insignificant compared to the other densities for it to 
contribute to the evolution in $H(z)$ in this redshift range. 

\begin{table*}[t]
\begin{center}
{\tiny
\caption{Optimized fits of $H(z)$}
\begin{tabular}{llllll}
&&& \\
\hline\hline
Model & Optimized parameters (conventional units) &  $\chi _{\rm dof}^2$ & Likelihoods && \\
\hline
$R_h=ct$ & $H_0=63.3\pm7.7$  & 0.566 & AIC: $82.9\%$ & KIC: $93.0\%$ & BIC: $94.3\%$ \\
$w$CDM & $H_0=68.0\pm7.9$, $\Omega_{\rm m}=0.31\pm0.16$, 
$w_{\rm de}=-0.91_{+0.42}^{-0.94}$ & 0.579 &AIC: $17.1\%$ & KIC: $\;\,7.0\%$ & BIC: $\;\,5.7\%$ \\
$\Lambda$CDM & $H_0=73.3\pm3.5$, $\Omega_{\rm m}=0.28\pm0.04$,
$\Omega_{\rm de}=0.60\pm0.13$ & 0.580 &AIC: $17.1\%$ & KIC: $\;\,7.0\%$ & BIC: $\;\,5.7\%$ \\
\hline\hline
\end{tabular}
}
\end{center}
\vskip 0.2in
\end{table*}

For each model, we optimize the fit by finding the set of parameters that
minimize the $\chi^2$, $H_0$ in the case of $R_{\rm h}=ct$; $H_0$,
$\Omega_{\rm m}$ and $w_{\rm de}$ in the case of $w$CDM; and $H_0$,
$\Omega_{\rm m}$ and $\Omega_{\rm de}$ for $\Lambda$CDM, as is
evident from Equations~(3) and (4). The results are summarized in Table 1,
and the corresponding best fits for $R_{\rm h}=ct$ and $w$CDM 
are shown in figure~1. These optimized
parameter values are quoted with one-sigma standard errors, calculated
from the corresponding $\chi^2$-distribution for each model. With $25-1=24$
degrees of freedom, the reduced $\chi_{\rm dof}^2$ for the $R_{\rm h}=ct$ 
model is 0.566. By comparison, the optimal $w$CDM fit has $25-3=22$ 
degrees of freedom, and a corresponding $\chi_{\rm dof}^2=0.579$. The
best-fit $\Lambda$CDM model is similar to this, with $\chi_{\rm dof}^2=0.58$. 
These fits suggest that $R_{\rm h}=ct$ is at least as good as $w$CDM/$\Lambda$CDM
in accounting for the model-independent $H(z)$ measurements, especially 
since it has only one free parameter. In particular, note how well it
accounts for the most recent $z\sim 2$ measurement; though, to be fair,
the error bar at this redshift is still quite large, so that $w$CDM/$\Lambda$CDM
cannot be excluded simply on the basis of this value alone.

On statistical grounds, however, the $R_{\rm h}=ct$ expansion rate versus
redshift is far more likely to be correct than that in $w$CDM/$\Lambda$CDM.
To compare the evidence for and against competing models, the use of
various information criteria is now fairly common in cosmology (see, e.g., 
Takeuchi 2000; Liddle 2004, 2007; Tan and Biswas 2012). For example, the Akaike
Information Criterion (AIC), 
\begin{equation}
{\rm AIC}=\chi^2+2f\;,
\end{equation}
provides an enhanced `goodness of fit,' extending the usual $\chi^2$ statistic 
by also taking into account the number of free parameters $f$ in each model. The
AIC ranks two or more competing models, and yields a numerical measure
of confidence that each model is preferred (Akaike 1973; Burnham and Anderson
2002, 2004). Clearly, the AIC prefers models with fewer parameters,
as long as the others do not provide a substantially better fit to the data.
Informally, in a one-on-one comparison between two models, model $\alpha$ has
a likelihood
\begin{equation}
L_{\alpha}={\exp(-{\rm AIC}_\alpha/2)\over \exp(-{\rm AIC}_1/2)+\exp(-{\rm AIC}_2/2)}
\end{equation}
of being the best choice. For the analysis reported here, these likelihoods are
$82.9\%$ for $R_{\rm h}=ct$ and $17.1\%$ for either $w$CDM or $\Lambda$CDM (see Table 1). 

An alternative to the AIC (Cavanaugh 1999) is known as the Kullback Information
Criterion (KIC), ${\rm KIC}=\chi^2+3f$,
which disfavors overfitting more than does the AIC. Not surprisingly, the
corresponding values of ${\rm KIC}=93.0\%$ for $R_{\rm h}=ct$ and
$7.0\%$ for either $w$CDM or $\Lambda$CDM, therefore favor the former by 
a greater amount.

A better-known alternative, known as the Bayes Information Criterion (BIC),
is actually not based on information theory, but rather on an asymptotic
approximation to the results of a conventional Bayesian inference procedure
for comparing two models (Schwarz 1978):
\begin{equation}
{\rm BIC}=\chi^2+f(\ln n)\;.
\end{equation}
The BIC suppresses overfitting very strongly when the number of data
points $n$ is large. In the sample used here, $n=25$, and we find that
${\rm BIC}=94.3\%$ for $R_{\rm h}=ct$, versus only $5.7\%$ for either 
$w$CDM or $\Lambda$CDM.

According to all three statistics, the expansion rate versus redshift
predicted by $R_{\rm h}=ct$ is more likely than that of $w$CDM/$\Lambda$CDM
to be closer to the `correct' cosmology. This consistency in the
results is quite important because, whereas the BIC is based on 
Bayesian statistics, the AIC and KIC are not. When $n$ is large,
any chosen Bayesian `priors' over the parameters of the
individual models drop out (Kuha 2004). However, the prior
distribution over the choice of models does not. In other words,
though the model distribution is often reasonably assumed to be `flat', 
i.e., to give each model an equal a priori likelihood,
a case can sometimes be made that one model is to be preferred a priori
over the other. 

The idea is that in the light of the data, e.g., the BIC computed 
from the data, the relative likelihood of the two models being compared 
is multiplied by a Bayes factor. When one assumes that the models are 
equally likely a priori, the multiplication by the Bayes factor yields 
(if the BIC is being used) their likelihoods being given by the 
BIC-derived Bayes weights, according to the usual formula. But the 
models don't necessarily need to be taken, a priori (i.e., pre-data),
to be equally likely.

So for the BIC one needs to assess beforehand whether
either $w$CDM/$\Lambda$CDM or $R_{\rm h}=ct$ is preferred.
Of course, this is precisely the problem with attempting
to prioritize models before testing them against the data.
Different workers can have different subjective opinions.
As of today, these two models have been tested against each
other using diverse types of data, all of which have thus
far favored $R_{\rm h}=ct$. Should one assign a prior
distribution for the models, the evidence now suggests
that $R_{\rm h}=ct$ is preferred. But to avoid such
biases based on subjective points of view, the only
sensible approach with the work reported in this paper 
is to simply let the data speak for themselves, and assume 
that the prior distribution over the choice of models is 
flat. 

The fact that we here use three different information criteria to
evaluate the models, of which the AIC and KIC have their own
foundations and do not belong to Bayesian statistics, strengthens
our overall conclusions. The inclusion of new high-$z$ measurements 
of $H(z)$ using cosmic chronometers has reinforced the results of 
cosmological model comparisons, all of which have thus far favored 
$R_{\rm h}=ct$ over $w$CDM/$\Lambda$CDM. 

Nonetheless, in a similar study of $H(z)$ measurements based on BAO 
observations, rather than cosmic chronometers, Shafer (2015) reaches 
a different conclusion. Unfortunately, Shafer's analysis is a good 
example of the unwitting use of model-dependent measurements to test 
competing cosmologies. We reiterate the obvious, though often forgotten, 
necessity of using only truly model-independent data to conduct 
statistically fair comparisons between different models. Without 
adequate explanation, Shafer opts to completely ignore measurements of 
$H(z)$ using cosmic chronometers, relying instead on BAO measurements, 
in spite of the fact that the deficiencies in the latter have already 
been discussed at length, e.g., in Melia \& Maier (2013). He appears to 
be unaware of the significant limitations of all but the most recent 
two BAO measurements at $z=0.57$ (Anderson et al. 2014) and $2.34$ 
(Delubac et al. 2015) for this kind of work. All previous applications 
of the galaxy two-point correlation function to measure a redshift-dependent 
scale were limited by the difficulty in disentangling the acoustic length 
in redshift space from redshift distortions due to internal gravitational 
effects (L\'opez-Corredoira 2014). With this process, one had to either 
pre-assume a particular model, or adopt prior parameter values to 
estimate the level of contamination. And the wide range of possible 
distortions for the same correlation-function shape resulted in seriously 
large errors.

To illustrate how significant these limitations are, and how biased the
use of BAO measurements of $H(z)$ can be, consider the considerable
disparity between Shafer's conclusion that these data strongly favor
$\Lambda$CDM over $R_{\rm h}=ct$, and the results of the 
Alcock-Paczy\'nski test using only the two most recent measurements
that avoid such model-dependent biasing (Melia \& L\'opez-Corredoira
2015). The much more precise determination of the Ly-$\alpha$ and 
quasar auto- and cross-correlation functions at $z=2.34$ has resulted in the 
measurement of BAO peak positions to better than $\sim 4\%$ accuracy
(Delubac et al. 2015). A similar accuracy has been achieved through the 
application of a technique of reconstruction to improve the signal/noise 
measurement of the BAO peak position in the anisotropic distribution 
of SDSS-III/BOSS DR11 galaxies at $\langle z\rangle=0.57$ (Anderson 
et al. 2014).

Unlike previous measurements, the actual shape of the BAO peak does 
not affect the calculation of its centroid position for these two
high-precision cases, both along the line-of-sight 
and in the direction perpendicular to it, when its FWHM is very narrow. The 
peak's narrowness mitigates the impact of redshift distortions, which affect 
the peak's amplitude, but not its location. Notice, for instance, that, although 
the errors for the redshift distortions quoted in Table 2 of Delubac et al. (2015)
are very large, the relative error bars for $H(z)$ are much smaller. In addition, 
the application of the Alcock-Paczy\'nski test to these two precision measurements
ensures that any model dependence in the BAO data is completely removed
because in this test, both the unknown acoustic scale and the Hubble constant $H_0$
completely cancel out, not to mention that this test is also completely independent
of any possible redshift evolution in the acoustic length. As shown by Melia \& 
L\'opez-Corredoira (2015), this test indicates that the concordance model is 
excluded at a $99.34\%$ C.L., while the probability that $R_{\rm h}=ct$ is 
consistent with these data is $\sim 96\%$. 

In other words, the true model-independent BAO data overwhelmingly 
favor $R_{\rm h}=ct$ over $w$CDM/$\Lambda$CDM, in sharp contrast to the 
conclusions drawn by Shafer (2015). There is therefore no merit to
his claim that the cosmic-chronometer measurements of $H(z)$, which
strongly favor $R_{\rm h}=ct$ over $\Lambda$CDM, should be 
supplanted by the BAO measurements, which he erroneously believes
yield the opposite result. 

Shafer (2015) also considers Type Ia SN data, and though SNe are not 
the subject of the present paper, it is nonetheless useful to point 
out the bias in this analysis as well. These issues have already been 
discussed and published in, e.g., Wei et al. (2015b), so we will provide
only a brief summary of the key issues here.

The statistical analysis of Type~Ia SNe can be improved with the merger 
of disparate sub-samples. For example, the Union2.1 catalog (Kowalski
et~al.\ 2008; Suzuki et~al.\ 2012), which currently includes $\approx580$
SN detections, offers several statistical advantages, but each 
sub-sample comes with its own set of systematic and intrinsic 
uncertainties. These are subsumed into unknown {\it intrinsic}
dispersions $\sigma_{\rm int}$'s (one for each sub-sample), which
makes it difficult to fit cosmological models. The commonly followed
approach, and apparently the one also followed by Shafer (2015), is
to minimize an overall~$\chi^2$, while constraining the $\chi^2_{\rm dof}$ 
of each sub-sample to equal unity. It is hardly surprising, then, that
the overall $\chi^2_{\rm dof}$ is very nearly one. However, a correct
statistical approach would estimate the unknown $\sigma_{\rm int}$'s 
simultaneously with all other parameters (Kim 2011; Wei et al. 2015b). 
For this, the use of maximum likelihood estimation (MLE) has been shown 
to yield superior results (D'Agostini 2005; Kim 2011, though the presence 
of multiple $\sigma_{\rm int}$'s complicates the analysis when the number of
merged sub-samples is greater than $\approx2-3$.

The Union2.1 catalog contains at least 17 sub-samples. In reality, therefore,
the total number of ``nuisance" parameters is ~20, since all of the
$\sigma_{\rm int}$'s need to be recalibrated for each independent
cosmological model in a truly unbiased test. This is unrealistic.
Note that this also means the expressions used by Shafer to estimate the
AIC and BIC are incorrect. Since the $\sigma_{\rm int}$'s themselves are 
not known a priori, the AIC and BIC must be calculated in terms of the
likelihood function, not $\chi^2$, as described in Wei et al. (2015b). 
In such cases, it is therefore preferable to work with a few large sub-samples, 
rather than many smaller ones. Fortunately, about half of the Type~Ia SNe in 
Union2.1 came from the single, homogeneous sample known as the SNLS 
(Guy et~al.\ 2010), and since the same instruments and reduction techniques 
were employed for all 252 of these high-$z$ ($0.15<z<1.1$) events, a single 
$\sigma_{\rm int}$ characterizes the unknown intrinsic scatter in this
homogeneous sample. Notice, in particular, that this sample spans the important
range of redshifts within which the transition from deceleration to acceleration
is thought to have occurred. 

This was the approach followed by Wei et al. (2015b), who concluded from the
careful analysis of the SNLS catalog that the BIC favors $R_{\rm h}=ct$
over $w$CDM/$\Lambda$CDM with a likelihood of $\sim 90\%$ versus only $\sim 10\%$.
In contrast, Shafer based his analysis on merged sub-samples, with the additional 
complication of unknown $\sigma_{\rm int}$'s and the use of questionable
statistical techniques. But he does not adequately explain why his result 
is completely reversed when one uses a single sample with uniform systematics. 
Surely the outcome of a detailed Type Ia SN analysis, if carried out properly,
should be robust enough to emerge intact regardless of whether one uses a single 
large sample, or many smaller sub-samples. 

Indeed, to illustrate this point, we (Melia et al. 2015b) carried out a similar 
analysis to that of Wei et al. (2015b), this time using both the SNLS 
and Sloan Digital Sky Survey (SDSS-II) events in the Joint Lightcurve Analysis 
(JLA) of Betoule et al. (2014). But a careful screening of these two
sub-samples shows that the SNLS and SDSS-II SNe have a small 
relative offset in their measured magnitudes, possibly due to slight 
differences in the background subtraction between the two surveys. A 
systematic shift such as this is not adequately handled by introducing
sub-sample-specific~$\sigma_{\rm int}$'s. Instead, a more meaningful approach
is to use a single sample-wide $\sigma_{\rm int}$, together with a systematic 
magnitude offset parameter $\Delta M_{\rm offset}$ between the SNLS and 
SDSS\nobreakdash-II events. The results of this study show that the JLA
merged SN sample favors $R_{\rm h}=ct$ over $w$CDM/$\Lambda$CDM with a likelihood 
of $\sim 88\%$ versus only $\sim 12\%$, in complete agreement with the 
previous study using solely the SNLS catalog on its own. 

This is the type of consistency one expects from the analysis of Type Ia SNe
if the analysis is being handled correctly and independently of any inherent
bias. In particular, one cannot use parameters optimized for one model
to test another, and the use of a single large sample should not completely
reverse the results of analysis of more complicated, merged catalogs. In
summary, Shafer's (2015) claim, that $w$CDM/$\Lambda$CDM is preferred over $R_{\rm h}=ct$
by the BAO-measured values of $H(z)$ and the analysis of Type Ia SNe, is
based on a combination of improper data selection and a statistically
flawed analysis. Instead, the results of the present paper show that 
model-independent measurements of $H(z)$ strongly favor $R_{\rm h}=ct$
over $w$CDM/$\Lambda$CDM, in complete agreement with most of the other 
published one-on-one comparative tests between these two cosmologies.

\section{Conclusions}
In reporting the high-$z$ measurements of $H(z)$
using cosmic chronometers, Moresco (2015) applied this diagnostic
to $\Lambda$CDM in order to demonstrate its power for improving 
the precision with which the model parameters may be optimized.
He found a detectable improvement ($\sim 5\%$) in the 
inferred value of $\Omega_{\rm m}$ and $w_{\rm de}$ over 
previous studies restricted to measurements of $H(z)$
at $z<1.75$.

In our previous study using a sample similarly restricted to 
$z\lesssim 1.75$ (Melia and Maier 2013), we derived
respective Bayesian posterior probabilities of $91.2\%$
for $R_{\rm h}=ct$ and $8.8\%$ for $w$CDM/$\Lambda$CDM. In the
work reported here, the addition of 4 new measurements at 
$z\lesssim 0.3$ by Zhang et al. (2014) and, especially, the 
two new measurements at $z\sim 2$ by Moresco (2015), have 
not only confirmed our earlier results, but have strengthened
the statistical significance of this important one-on-one
comparison between competing cosmological models. We have
found that the model-independent cosmic chronometer data
prefer the $R_{\rm h}=ct$ Universe over $w$CDM/$\Lambda$CDM with
a BIC likelihood of $\sim 95\%$ versus only $\sim 5\%$.

We therefore confirm Moresco's (2015) conclusion that 
upcoming, additional measurements of $H(z)$ with expanded 
surveys at high redshifts (e.g., Laureijs et al. 2011)
will provide us with one of the most powerful probes of 
the cosmic spacetime in the local Universe.

In so doing, we have achieved the primary goal of this
work---to independently confirm the outcome of previous
comparative tests between $R_{\rm h}=ct$ and $w$CDM/$\Lambda$CDM
using SNeIa and BAO data within the critical redshift 
range ($0\lesssim z \lesssim 2$) where the transition 
from decelerated to accelerated expansion is thought 
to have occurred. 

These three diagnostics, the SNeIa Hubble diagram (Melia et al.
2015b), the Alcock-Paczy\'nski BAO test (Melia and L\'opez-Corredoira
2015), and now the $H(z)$-redshift relation, each probes the 
cosmic expansion in its own unique way. The fact 
that all three have self-consistently demonstrated that 
$R_{\rm h}=ct$ is preferred over $w$CDM/$\Lambda$CDM with comparable 
statistical significance argues strongly in favor of the
zero active mass equation-of-state. Our results affirm
the influence of dark energy in the cosmic fluid; but
not in the guise of a cosmological constant.

\acknowledgments
F.M. is grateful to Amherst College for its support through
a John Woodruff Simpson Lectureship, and to Purple Mountain Observatory in Nanjing,
China, for its hospitality while part of this work was being carried out. This work
was partially supported by grant 2012T1J0011 from The Chinese Academy of Sciences
Visiting Professorships for Senior International Scientists.

\end{document}